\documentclass{llncs}

\ifx\pdfoutput\undefined
\usepackage{graphicx}
\usepackage[dvips]{epsfig}
\else
\usepackage[pdftex]{graphicx}
\fi
\usepackage{amsmath}
\usepackage{array}
\usepackage{latexsym}
\def\HS{\hbox{$\hbox{S}_H$}}
\def\HA{\hbox{$\hbox{A}_H$}}
\def\HF{\hbox{$\hbox{F}_H$}}
\def\HIS{\hbox{$\hbox{S}_{{H\!I}}$}}
\def\HIA{\hbox{$\hbox{A}_{{H\!I}}$}}
\def\HIF{\hbox{$\hbox{F}_{{H\!I}}$}}
\def\BS{\hbox{$\hbox{S}_B$}}
\def\BA{\hbox{$\hbox{S}_B$}}
\def\BF{\hbox{$\hbox{S}_B$}}

\begin{document}
\frontmatter          % for the preliminaries
\pagestyle{headings}  % switches on printing of running heads

\mainmatter

\title{Software Assumptions Failure Tolerance:\\Role, Strategies, and Visions}

\author{Vincenzo De Florio}
\institute{University of Antwerp\\
	Department of Mathematics and Computer Science\\
	Performance Analysis of Telecommunication Systems group\\
	Middelheimlaan 1, 2020 Antwerp, Belgium. \vspace*{3pt} \\
%\and
	Interdisciplinary Institute for Broadband Technology\\
	Gaston Crommenlaan 8, 9050 Ghent-Ledeberg, Belgium.}

\maketitle

\begin{abstract}
At our behest or otherwise, while our software is being
executed, a huge variety of design assumptions is 
continuously matched with the truth of the current condition. While 
standards and tools exist to express and verify some of these 
assumptions, in practice most of them end up being either sifted off or 
hidden between the lines of our codes. Across the system layers, a 
complex and at times obscure web of assumptions determines the 
quality of the match of our software with its deployment platforms and 
run-time environments. Our position is that it becomes increasingly 
important being able to design software systems with architectural and 
structuring techniques that allow software to be decomposed to reduce its complexity,
but without hiding in the process vital hypotheses and assumptions.
In this paper we discuss this problem, introduce
three potentially dangerous consequences of its denial,
and propose three strategies to facilitate their treatment. 
Finally we propose our vision towards a new holistic approach
to software development to overcome the shortcomings offered
by fragmented views to the problem of assumption failures.
\end{abstract}

%%%%%%%%%%%%%%%%%%%%%% I N T R O D U C T I O N %%%%%%%%%%%%%%%%%%%%%%%%%%
%%%%%%%%%%%%%%%%%%%%%%%%%%%%%%%%%%%%%%%%%%%%%%%%%%%%%%%%%%%%%%%%%%%%%%%%%
%%
%%{{{  Intro
%%%%%%%%%%%%%%%%%%%%%%%%%%%%%%%%%%%%%
\section{Introduction}\label{s:intro}
%%%%%%%%%%%%%%%%%%%%%%%%%%%%%%%%%%%%%
We are living in a society that cannot do without 
computer systems. Services supplied by computer systems 
have permeated our environments and deeply changed our 
societies and the way we live in them. Computers pervade 
our lives, integrating themselves in all environments. At 
first confined in large control rooms, now they take the 
form of tiny embedded systems soon to be ``sprayed'' on 
physical entities so as to augment them with advanced 
processing and communication capabilities. Thus it is 
very much evident to what extent we depend on computers.  
What is often overlooked by many is the fact 
that most of the logics behind computer services 
supporting and sustaining our societies lies in the 
software layers. Software has become the point of 
accumulation of a large amount of complexity~\cite{Sch06}. It is 
ubiquitous, mobile, and has pervaded all aspects of our 
lives. What is more important for this discussion, software
is the main culprit behind the majority of 
computer failures~\cite{Lyu98a,Lyu98b,Lapr98}.

Among the reasons that brought to this state of things we 
focus our attention here on a particular one. Clever 
organizations and system structures allowed the 
visible complexity of software development to be reduced---at first 
through modules and layers, then by means of objects, and 
more recently with services, components, aspects, and models. As 
a result, we have been given tools to compose and 
orchestrate complex, powerful, and flexible 
software-intensive systems in a relatively short amount 
of time. The inherently larger flexibility of software 
development turned software into the ideal ``location'' 
where to store the bulk of the complexity of nowadays' 
computer-based services. Unfortunately, this very same 
characteristic of software makes it also considerably 
\emph{fragile to changes}~\cite{Sch06}. In particular
software's flexibility 
also means that most of the assumptions drawn at 
design-time may get invalidated when the software system 
is ported, reused, redeployed, or simply when it is 
executed in a physical environment other than the 
one originally meant for. This means that 
truly resilient software systems demand special care to 
\emph{assumption failures detection}, \emph{avoidance}, 
and \emph{recovery}. Despite this fact, 
no systematic approach allows yet for the expression 
and verification of hypotheses regarding the expected 
properties and behaviors of

\begin{itemize} 
  \item the hardware components (e.g. the failure semantics 
        of the memory modules we depend on);
  \item third-party software (e.g. the reliability of an open-source 
        software library we make use of);
  \item the execution environment (e.g. the security provisions 
        offered by the Java execution environment we are currently using);
  \item the physical environment (e.g., the characteristics of the faults
        experienced in
        a space-borne vehicle orbiting around the sun). 
\end{itemize}

While several tools exist, in practice
most of the above assumptions often end up being either 
sifted off or ``hardwired'' in the executable code. As 
such, those removed or concealed hypotheses cannot be 
easily inspected, verified, or maintained.
%\footnote{TMP:
            %Model Driven Engineering, UML etc are effective tools...
            %but often shift the problem...often we end up still ``wrestling''
            %with the code~\cite{Sch06}...}.
Despite the availability of several conceptual and
practical tools---a few examples of which are briefly discussed
in Sect.~\ref{s:reltech}---still
we are lacking methodologies and architectures to tackle this 
problem in its complex entirety---from design-time to the
various aspects of the run-time. As a consequence, our software systems 
often end up being entities whose structure, properties, 
and dependencies are not completely known, hence at times
deviate from their intended goals.

Across the system layers, a complex and at times obscure 
``web'' of software machines is being executed 
concurrently by our computers. Their mutual dependencies 
determine the quality of the match of our software with 
its deployment platform(s) and run-time environment(s) 
and, consequently, their performance, cost, and in general
their quality of service and experience. At our behest or 
otherwise, a huge variety of design assumptions is 
continuously matched with the truth of the current 
conditions. A hardware component assumed to be available; 
an expected feature in an OSGi bundle or in a web browser 
platform; a memory management policy supported by 
a mobile platform~\cite{mw09.12}, or ranges of operational 
conditions taken for granted at all times---all are but 
assumptions and all have a dynamically varying truth 
value.

Our societies, our very lives, are often entrusted to 
machines driven by software; weird as it may sound, 
in some cases this 
is done without question---as an act of faith as it 
were. This is clearly unacceptable. The more we rely on 
computer systems---the more we depend on their correct 
functioning for our welfare, health, and economy---the 
more it becomes important to design those systems with 
architectural and structuring techniques that allow 
software complexity to be decomposed, but without hiding 
in the process those hypotheses and assumptions 
pertaining e.g. the target execution environment and the 
expected fault- and system models.

Our position is that existing tools will have to be 
augmented so as to minimize the risks of assumption 
failures e.g. when porting, deploying, or moving software 
to a new machine. We envision novel autonomic run-time 
executives that continuously verify those hypotheses and 
assumptions by matching them with endogenous knowledge 
deducted from the processing subsystems as well as 
exogenous knowledge derived from their execution and 
physical environments. Mechanisms for propagating such
knowledge through all stages of software development would
allow the chances of assumptions failures to be considerably reduced.
The ultimate result we envisage is the ability to express 
truly assumption failure-tolerant software systems, i.e., 
software systems that endorse provisions to efficiently 
and effectively tackle---to some agreed upon extent---the 
problem of assumption failures.

This paper makes three main contributions.
A first one is exposing our 
vision of assumption failure-tolerant software systems. 
Such systems explicitly address three main ``hazards'' of 
software development, which we call the Horning 
syndrome, the Hidden Intelligence syndrome, and the 
Boulding syndrome. Assumption failures and the three 
syndromes are presented in Sect.~\ref{s:hazards}. 
A second contribution is introducing the
concept of assumption failure-tolerant software systems
and providing three examples of strategies---one for
each of the above syndromes.
This is done in Sect.~\ref{s:aftss}.
A third contribution is our vision of a holistic approach
to resilient software development, where the concept of
assumption failure plays a pivotal role. Such vision---introduced
after a brief overview of related and complementary
technologies in Sect.~\ref{s:reltech}---is the subject
of Sect.~\ref{s:vision}.
The paper is concluded by Sect.~\ref{s:end} in which we
summarize our main lessons learned and provide our conclusions.

%%%%%%%%%%%%%%%%%%%%%%%%%%%%%%%%%%%%%%%%%%%%%%%%%%%%%%%%%%%%%%%% 
\section{Three Hazards of Software Development}\label{s:hazards} 
%%%%%%%%%%%%%%%%%%%%%%%%%%%%%%%%%%%%%%%%%%%%%%%%%%%%%%%%%%%%%%%% 

As mentioned before, assumption failures may have dire 
consequences on software dependability. In what follows we 
consider two well known exemplary cases from which we 
derive a base of three ``syndromes'' that we 
deem as the main hazards of
assumption failures. We assume the reader to be already 
familiar with the basic facts of those two cases. Furthermore, we shall 
focus our attention only on a few aspects and 
causes---namely those more closely related to the subject at hand.

\subsection{Case 1: Ariane 5 Flight 501 Failure}
On June 4, 1996, the maiden flight of the Ariane 5 rocket 
ended in a failure just forty seconds after its lift-off.  
At an altitude of about 3,700 meters, the launcher veered 
off its flight path, broke up and exploded.  After the 
failure, the European Space Agency set up an independent 
Inquiry Board to identify the causes of the failure. The 
Inquiry Board unravelled several reasons, the most 
important of which was a failure in the so-called 
Inertial Reference System (IRS), a key component 
responsible for flight attitude and movement control in 
space. Being so critical for the success of the mission, 
the IRS adopted a simple hardware fault-tolerance 
design pattern: two identical replicas were operating in 
parallel (hot standby), executing the same software system. As 
mentioned before, we shall not focus here on all the 
design faults of this scheme, e.g. its lack of design 
diversity~\cite{Avi85}. Our focus will be on one of 
the several concomitant causes, namely a software reuse error 
in the IRS. The Ariane 5 software included software 
modules that were originally developed and successfully 
used in the Ariane 4 program. Such software was written 
with a specific physical environment as its reference. 
Such reference environment was characterized by well 
defined ranges for several flight trajectory parameters. 
One such parameter was the rocket's maximum horizontal 
velocity. In the Ariane 4, horizontal velocity could be 
represented as a 16-bit signed integer. The Ariane 5 was 
a new generation, thus it was faster. In particular 
horizontal velocity could not be represented in a signed 
short integer, which caused an overflow in both IRS 
replicas. This event triggered a chain of failures that 
led the rocket to complete loss of guidance and attitude 
information shortly after the start of the ignition 
sequence. Now completely blind and unaware, the Ariane 5 
committed self destruction as an ultimate means to 
prevent any further catastrophic failures.

The Ariane 5 failure provides us with several lessons---in the 
rest of this subsection we shall focus on two of them.

\subsubsection{Horning Syndrome.}
The Ariane 5 failure warns us of the fact that an assumption 
regarding the target physical environment of a software 
component may clash with a real life fact. In the case at 
hand, the target physical environment was assumed to be 
one where horizontal velocity would not exceed some 
agreed upon threshold. This assumption clashed with the 
characteristics of a new target environment.

The term we shall use to describe this event is 
``assumption failure'' or ``assumption-versus-context 
clash''. The key lesson in this case is then that the 
physical environment can play a fundamental role in 
determining software quality.
By paraphrasing a famous quote by Whorf, the environment
shapes the way our fault-tolerance software is constructed
and determines how dependable it will ultimately be.
James Horning described 
this concept through his well known quote~\cite{Hor98}:

\begin{quote}
``What is the most often overlooked risk in 
software engineering?\\
That the environment will do something the 
designer never anticipated.''
\end{quote}

This is precisely what happened in the case of the 
failure of the Ariane 5's IRS: new unanticipated 
environmental conditions violated some design 
assumptions. For this reason we call this class of 
assumption failures hazards ``the Horning Syndrome'', or 
\HS{} for brevity. For the same reason we shall use the 
terms ``Horning Assumptions'' (\HA) and ``Horning 
Failures'' (\HF) respectively to refer to this class of 
assumptions and of failures.

In what follows we shall use lowercase letters in Italics to denote 
assumptions. Given a letter representing an assumption, the same letter 
in bold typeface shall
represent the true value for that assumption. As an 
example, the Ariane-5 failure was caused (among other reasons) by a 
clash between $f$: \{``Horizontal Velocity can be 
represented by a short integer''\} and \textbf{f}: 
\{``Horizontal velocity is now $n$''\}, 
where $n$ is larger than the maximum short integer.

\subsubsection{Hidden Intelligence Syndrome.}
The second aspect we deem important to highlight in the 
context of the failure of the IRS is related to a lack of 
propagation of knowledge. The Horning Assumption that led 
to this Horning Failure originated at Ariane 4's design 
time. On the other hand the software code that 
implemented the Ariane 4 design did not include any 
mechanism to store, inspect, or validate such assumption. 
This vital piece of information was simply lost. This 
loss of information made it more difficult to verify the 
inadequacy of the Ariane 4 software to the new 
environment it had been deployed. We call an accident 
such as this a case of the Hidden Intelligence Syndrome 
(\HIS). Consequently we use the terms Hidden Intelligence 
Assumption (\HIA) and Hidden Intelligence Failure (\HIF).

Unfortunately accidents due to the \HS{} and the \HIS{} 
are far from being uncommon---computer history is 
crowded with examples, with a whole range of 
consequences. In what follows we highlight this fact in 
another well known case---the deadly Therac-25 failures.

%%%%%%%%%%%%%%%%%%%%%%%%%%%%%%%%%%%%%%%%%%%%
\subsection{Case 2: The Therac-25 Accidents}
%%%%%%%%%%%%%%%%%%%%%%%%%%%%%%%%%%%%%%%%%%%%
The Therac-25 accidents have been branded as ``the most 
serious computer-related accidents to 
date''~\cite{Lev95}. Several texts describe and analyze 
them in detail---including the just cited one. As we did 
for the Ariane 5, here we shall not provide yet another 
summary of the case; rather, we shall highlight the 
reasons why the Therac-25 is also a case of the 
above assumption hazards and of a third class of 
hazards.

The Therac-25 was a so-called ``linac,'' that is, a 
medical linear accelerator that uses accelerated 
electrons to create high-energy beams to destroy 
tumors with minimal impact on the surrounding healthy 
tissue. It was the latest member of a successful family 
of linacs, which included the Therac-6 and the 
Therac-20. Compared to its predecessors, model 25 was 
more compact, cheaper and had more functional features. In 
particular the cheaper cost was a result of several 
modifications including a substantial redesign of the 
embedded hardware-software platform. In the redesign, 
some expensive hardware services were taken over by the 
software layer. For instance it was decided to remove 
hardware interlocks that would shut the machine down in 
the face of certain exceptions.

There is evidence that several such exceptions had 
occurred while previous models, e.g. the Therac-20, were 
operative. Unfortunately, none of these occurrences were 
reported or fed back to the design process of the 
Therac-25. Had it been otherwise, they would have revealed that 
certain rare combinations of events triggered the 
emission of extraordinary high levels of energy 
beams---were it not for the safety interlocks present in 
the old models. History repeated itself with model 25, 
only this time the killer doses of beams \emph{were\/} 
emitted, resulting in the killing or serious injuring of 
several people.

\subsubsection{Another Case of the Horning Syndrome.} We 
observe how the Therac may be considered as a special 
case of Horning Assumption failure in which the 
``unanticipated behavior'' is due to endogenous causes 
and Horning's ``environment'' is the hardware platform. 
The ``culprit'' in this case is the clash between two 
design assumptions and two indisputable facts. 
Assumptions were fault assumption $f$: 
$\{$``\emph{No residual fault exists}''$\}$ and hardware 
component assumption $p$: $\{$``\emph{All exceptions are 
caught by the hardware and the execution environment, and 
result in shutting the machine down}''$\}$. The 
corresponding facts were $\mathbf{f}$: 
$\{$``\emph{Residual faults still exist}''$\}$, that is 
$\neg f$, and $\mathbf{p}$: $\{$``\emph{Exceptions exist 
that are not caught}''$\}$---that is, $\neg p$. The 
unanticipated behavior is in this case the machine still 
remaining operative in a faulty state, thus the violation 
of the safety mission requirements.

\subsubsection{Another Case of Hidden Intelligence.} As 
mentioned already, because of the failure-free behavior 
of the Therac-20, its software was considered as 
fault-free. Reusing that software on the new machine 
model produced a clash. Thus we could say that, for the 
Therac family of machines, a hardware fault-masking 
scheme translated into software hidden 
intelligence---that is, a case of the \HIS. Such hidden 
intelligence made it more difficult to verify the 
inadequacy of the new platform to its
operational specifications.

\subsubsection{Boulding Syndrome.} Finally we observe how the 
Therac-25 software, despite its real-time design goals, was 
basically structured as a quasi closed-world system. Such 
systems are among the naivest classes of systems in 
Kenneth Boulding's famous classification~\cite{Bou56}: 
quoting from the cited article, they belong to 
the categories of ``Clockworks'' (``simple dynamic system 
with predetermined, necessary motions'') and 
``Thermostats'' (``control mechanisms in which [\dots\kern-1.5pt] the 
system will move to the maintenance of any given 
equilibrium, \emph{within limits}'').  Such systems are 
characterized by predefined assumptions about their 
platform, their internal state, and the environment they 
are meant to be deployed in. They are closed, ``blind'' 
entities so to say, built from synchronous assumptions, 
and designed so as to be plugged in well defined hardware 
systems and environments whose changes, idiosyncrasies, 
or fluctuations most of them deliberately ignore. Using a 
well known English vernacular, they are ``\emph{sitting ducks}'' 
to change---they keep on doing their prescribed task, as 
defined at design time, irrespective of environmental 
conditions; that is, they lack the ability to detect and 
respond to deployment- and run-time changes.

Clearly the Therac machines and their software comply to 
this definition. In particular those machines were 
missing introspection mechanisms (for instance, 
self-tests) able to verify whether the target platform did 
include the expected mechanisms and behaviors.

A case like the Therac's---that is, when a clash exists 
between a system's Boulding category and the actual 
characteristics of its operational environment---shall be 
referred to in what follows as a case of the Boulding 
Syndrome (\BS). The above mentioned Boulding categories 
and clashes will also be respectively referred to as 
Boulding Assumptions (\BA) and Boulding Failures (\BF).

%%%%%%%%%%%%%%%%%%%%%%%%%%%%%%%%%%%%
\subsection{Preliminary Conclusions}
%%%%%%%%%%%%%%%%%%%%%%%%%%%%%%%%%%%%
By means of two well known cases we have shown how 
computer system failures can be the result of software 
assumption failures. Moreover, in so doing we have introduced 
three major hazards or syndromes requiring particular 
attention:

\begin{description}
 \item[Horning syndrome:] mistakenly not considering that the
   physical environment may change and produce unprecedented or
   unanticipated conditions;
 \item[Hidden Intelligence syndrome:] mistakenly concealing or
   discarding important knowledge for the sake of hiding
   complexity;
 \item[Boulding syndrome:] mistakenly designing a system with
   insufficient context-awareness with respect to the current
   environments.
\end{description}

In what follows we describe examples of strategies to
treat some cases of the three syndromes so as to decrease 
the risk to trigger assumption failures.

%%%%%%%%%%%%%%%%%%%%%%%%%%%%%%%%%%%%%%%%%%%%%%%%%%%%%%%%%%%%%%%%%%%%%
\section{Assumption Failure-Tolerant Software Systems}\label{s:aftss}
%%%%%%%%%%%%%%%%%%%%%%%%%%%%%%%%%%%%%%%%%%%%%%%%%%%%%%%%%%%%%%%%%%%%%

The key strategy we adopt here is to offer the designer the 
possibility to postpone the choice of one out of multiple alternative 
design-time assumptions to a proper future time 
(compile-time, deployment-time, run-time, etc.) In what 
follows we shall describe how to do so for the following 
classes of assumptions:

\begin{itemize} 
  \item Assumptions related to the failure semantics of hardware components.
  \item Assumptions related to the fault-tolerance design
        patterns to adopt.
  \item Assumptions related to dimensioning of resources.
\end{itemize}

\subsection{Assumptions on Hardware Components'
            Failure Semantics}
As we have already remarked, software depends on certain 
behaviors expected from the underlying hardware 
architecture. Hardware neutrality and the principles of 
layered design dictate that most of the actual processes 
and actors in the bare machine are not disclosed. Thus 
for instance we rarely know (and often care about) the 
particular technology of the main memory integrated 
circuits our software is making use of. 

This is a case of 
the Hidden Intelligence syndrome. 
By not expressing explicitly our requirements concerning the way hardware 
(e.g., memory modules) should behave
we leave the door open to 
dependability assumption failures.

As an example, while yesterday's software was running 
atop CMOS chips, today a common choice e.g. for airborne 
applications is SDRAM---because of speed, cost, weight, 
power and simplicity of design~\cite{Lad02}. But CMOS 
memories mostly experience single bit 
errors~\cite{1981coae.conf...66O}, while SDRAM chips are 
known to be subjected to several classes of severe 
faults, including so-called ``single-event 
effects''~\cite{Lad02}, i.e., 
a threat that can lead to total loss of a
whole chip. 
Examples include:

\begin{enumerate}
 \item Single-event latch-up (SEL),
   a threat that can bring to the loss of all data stored on 
   chip~\cite{WikiLatchup}.
 \item Single-event upset (SEU), leading to 
   frequent soft errors~\cite{WikiSoftError,WikiSEU}. 
 \item Single-event functional interrupt (SFI), i.e. a special 
   case of SEU that places the 
   device into a test mode, halt, or undefined state. The SFI
   halts normal operations, and requires
   a power reset to recover~\cite{Holbert}. 
\end{enumerate} 

Furthermore~\cite{Lad02} remarks 
how even \emph{from lot to lot\/} error and failure rates 
can vary more than one order of magnitude. In other 
words, the superior performance of the new generation of 
memories is paid with a higher instability and a trickier 
failure semantics.

\begin{figure}[t]
\centerline{\includegraphics[angle=90,width=0.5\textwidth]{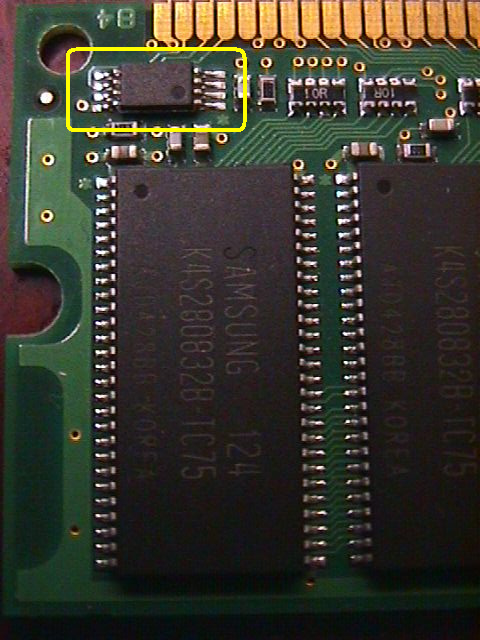}}
%\centerline{\includegraphics{MVC-008F.png}}
\caption{The 
Serial Presence Detect (yellow circle) allows information
about a computer's memory module, e.g. its manufacturer, model, size, 
and speed, to be accessed.}
\label{f:spd}
\end{figure}

Let us suppose for the time being that the software 
system at hand needs to be compiled
in order to be executed on the target 
platform. The solution we propose 
to alleviate this problem is as follows:

\begin{itemize}
\item First, we assume memory access is abstracted (for instance
through services, libraries, overloaded operators, or aspects).
This allows the actual memory access methods
to be specified in a second moment.
\item Secondly, a number of design-time hypotheses regarding the failure semantics
of the hardware memory subsystem are drawn. These may take the form
of fault/failure assumptions such as for instance:
  \begin{description}
  \item{$f_0$:} ``Memory is stable and unaffected by failures''.
  \item{$f_1$:} ``Memory is affected by transient faults and CMOS-like
		 failure behaviors''.
  \item{$f_2$:} ``Memory is affected by permanent stuck-at faults and CMOS-like
		failure behaviors''.
  \item{$f_3$:} ``Memory is affected by transient faults and SDRAM-like
		 failure behaviors, including SEL''.
  \item{$f_4$:} ``Memory is affected by transient faults and SDRAM-like
		 failure behaviors, including SEL and SEU''.
  \end{description}
\item For each assumption $f_i$ (in this case $0\le i\le4$) a diverse set of
memory access methods, $M_i$, is designed. With the exception of $M_0$, 
each $M_i$ is a fault-tolerant version specifically designed to
tolerate the memory modules' failure modes assumed in $f_i$.
\item To compile the code on the target platform, an Autoconf-like
toolset~\cite{Cal10} is assumed to be available. Special checking rules are coded
in the toolset making use of e.g. Serial Presence Detect (see Fig.~\ref{f:spd}) to
get access to information related to the memory modules on the target computer.
For instance, Linux tools such as ``\texttt{lshw}'' provide higher-level access to information
such as the memory modules' manufacturer, models, and characteristics
(see an example in Fig.~\ref{f:lshw}).
Such rules could access local or remote, shared databases reporting
known failure behaviors for models and even specific lots thereof.
Once the most probable memory behavior \textbf{f} is retrieved,
a method $M_j$ is selected
to actually access memory on the target computer.
Selection is done as follows: first we isolate
those methods that are able to tolerate \textbf{f}, then we
arrange them into a list ordered according to some cost function (e.g.
proportional to the expenditure of resources);
finally we select the minimum element of that list.
\end{itemize}

\begin{figure}[t]
\begin{verbatim}
     *-memory
          description: System Memory
          physical id: 1000
          slot: System board or motherboard
          size: 1536MiB
        *-bank:0
             description: DIMM DDR Synchronous 533 MHz (1.9 ns)
             vendor: CE00000000000000
             physical id: 0
             serial: F504F679
             slot: DIMM_A
             size: 1GiB
             width: 64 bits
             clock: 533MHz (1.9ns)
        *-bank:1
             description: DIMM DDR Synchronous 667 MHz (1.5 ns)
             vendor: CE00000000000000
             physical id: 1
             serial: F33DD2FD
             slot: DIMM_B
             size: 512MiB
             width: 64 bits
             clock: 667MHz (1.5ns)
\end{verbatim}
\caption{Excerpt from the output of command-line \texttt{sudo lshw}
on a Dell Inspiron 6000 laptop.}\label{f:lshw}
\end{figure}

The above strategy allows the designer to postpone the 
choice between alternative design-time assumptions to the 
right moment, that it, when the code is compiled on the 
chosen target action. A similar strategy could be 
embedded in the execution environment, e.g. a Web browser 
or a Java Virtual Machine. Such strategy could 
selectively provide access at deployment time to 
knowledge necessary to choose which of the design-time 
alternative assumptions has the highest chance to match 
reality. Note that our strategy helps avoiding \HIS{}
and brings the designer to explicitly deal with the problem
of assumption failures. Furthermore this is done with full
separation of the design concerns.

\subsubsection{Comparison with existing strategy.} 
A somewhat similar strategy is used for performance 
enhancement. Applications such as the mplayer video 
player~\cite{Mplayer} can take advantage of predefined knowledge about 
the possible target processor and enable optimized 
methods to perform some of their tasks. Mplayer declares 
this by displaying messages such as ``Using SSE optimized 
IMDCT transform'' or ``Using MMX optimized resampler''. 
Our procedure differs considerably from the mplayer's, as 
it focuses on non-functional (dependability) enhancements. Furthermore, it 
is a more general design methodology and makes use of 
knowledge bases. Meta-object protocols, compiler 
technology, and aspects could provide alternative way to 
offer similar services.

%%%%%%%%%%%%%%%%%%%%%%%%%%%%%%%%%%%%%%%%%%%%%%%%%%%%%%%%%%%%%%%%%%%%%%%%
\subsection{Choice of Fault-tolerance Design Patterns} \label{s:pattern}
%%%%%%%%%%%%%%%%%%%%%%%%%%%%%%%%%%%%%%%%%%%%%%%%%%%%%%%%%%%%%%%%%%%%%%%%
The choice of which design pattern to use is known to 
have a direct influence on a program's overall complexity 
and performance. What is sometimes overlooked is the fact 
that fault-tolerance design patterns have a strong 
influence on a program's actual ability to tolerate 
faults. For instance, a choice like the \textbf{redoing} 
design pattern~\cite{SKS96}---i.e., repeat on failure---\emph{implies\/}
assumption $e_1:$ \{``The 
physical environment shall exhibit transient faults''\}, 
while a design pattern such as \textbf{reconfiguration}---that is,
replace on failure---is
the natural choice after an assumption such as $e_2:$ 
\{``The physical environment shall exhibit permanent 
faults''\}. Of course clashes are always possible, which 
means in this case that there is a non-zero probability 
of a Horning Assumption failure---that is, a case of the 
\HS. Let us observe that:
%qual'e' il tasto di mamma? E lui batte due volte: mm

\begin{enumerate}
\item A clash of assumption $e_1$
  implies a livelock (endless repetition) as a result of redoing actions
  in the face of permanent faults.
\item A clash of assumption $e_2$ implies an unnecessary
  expenditure of resources as a result of applying
  reconfiguration in the face of transient faults.
\end{enumerate}

The strategy we suggest to tackle this case
is to offer the designer the possibility to
postpone the binding of the actual
fault-tolerance design pattern and to condition it
to the actual behavior of the environment.

\begin{figure}[t]
\centerline{\includegraphics[width=0.7\textwidth]{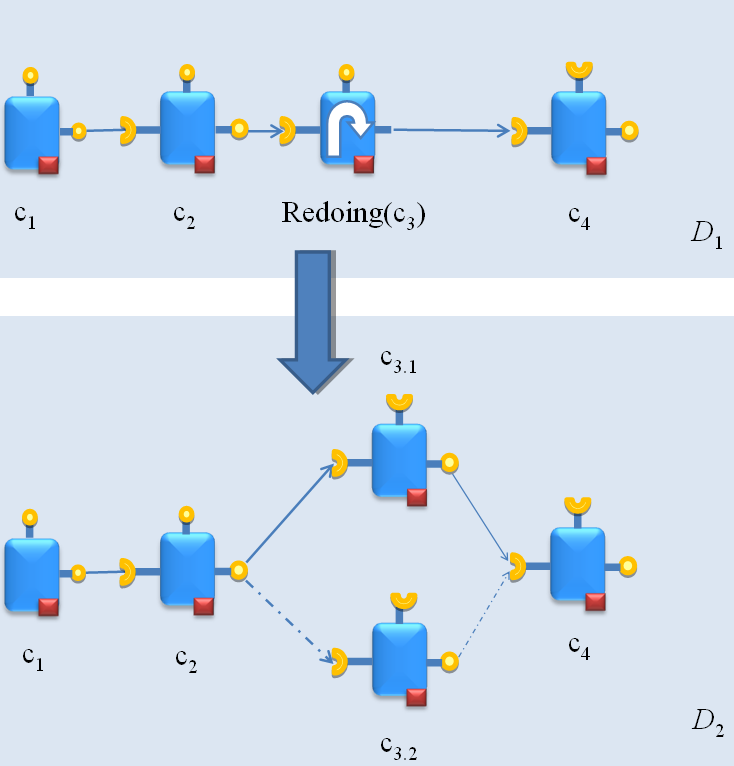}}
\caption{Transition from a redoing scheme $(D1)$ to
         a reconfiguration scheme $(D_2)$ is obtained by replacing
	 component $c_3$, which tolerates transient faults by redoing its computation,
         with a 2-version scheme where a primary component ($c_{3.1}$) is
         taken over by a secondary one ($c_{3.2}$) in case of permanent faults.}
\label{f:d1tod2}
\end{figure}

In what follows we describe a possible implementation
of this run-time strategy.
\begin{itemize}
\item First, we assume the software system
to be structured in such a way as to allow
an easy reconfiguration of its components.
Natural choices for this are service-oriented and/or
component-oriented architectures.
Furthermore we assume that the software architecture can be
adapted by changing a reflective meta-structure in the form
of a directed acyclic graph (DAG). A middleware supporting
this is e.g. ACCADA~\cite{GuD10+}.

\item Secondly, the designer draws a number of alternative
hypotheses regarding the faults to be experienced
in the target environments.
A possible choice could be for instance
$e_0$: ``No faults shall be experienced'' and then
$e_1$ and $e_2$ from above.
\item For each fault-tolerance assumption (in this case $e_1$
and $e_2$) a matching fault-tolerant design pattern is designed
and exported e.g. in the service or component registry.
The corresponding DAG snapshots are stored in
data structures $D_1$ and $D_2$.

\item Through e.g. publish/subscribe, the
supporting middleware component receives notifications
regarding the faults being detected by the main
components of the software system.
Such notifications are fed into an Alpha-count
filter~\cite{Bon96,BCDG00}, that is, a count-and-threshold
mechanism to discriminate between different types of
faults.
\item Depending on the assessment of the Alpha-count oracle,
either $D_1$ or $D_2$ are injected on the reflective DAG.
This has the effect or reshaping the software architecture
as in Fig.~\ref{f:d1tod2}.
Under the hypothesis of a correct oracle, such scheme avoids
clashes: always the most appropriate design 
pattern is used in the face of certain classes of faults.
\end{itemize}

The above strategy is a second example of a way
to postpone the 
choice among alternative design-time assumptions to the 
right moment---in this case at run-time, when the
physical environment changes its characteristics or when
the software is moved to a new and different environment.
As a consequence, our strategy has the effect to help
avoiding \HS{} and to force the designer not to
neglect the problem of assumption failures.

\begin{figure}[t]
%\hspace*{-0.05\textwidth}\includegraphics[width=1.1\textwidth]{alphacount1.png}
\hspace*{-0.05\textwidth}\includegraphics[width=1.1\textwidth]{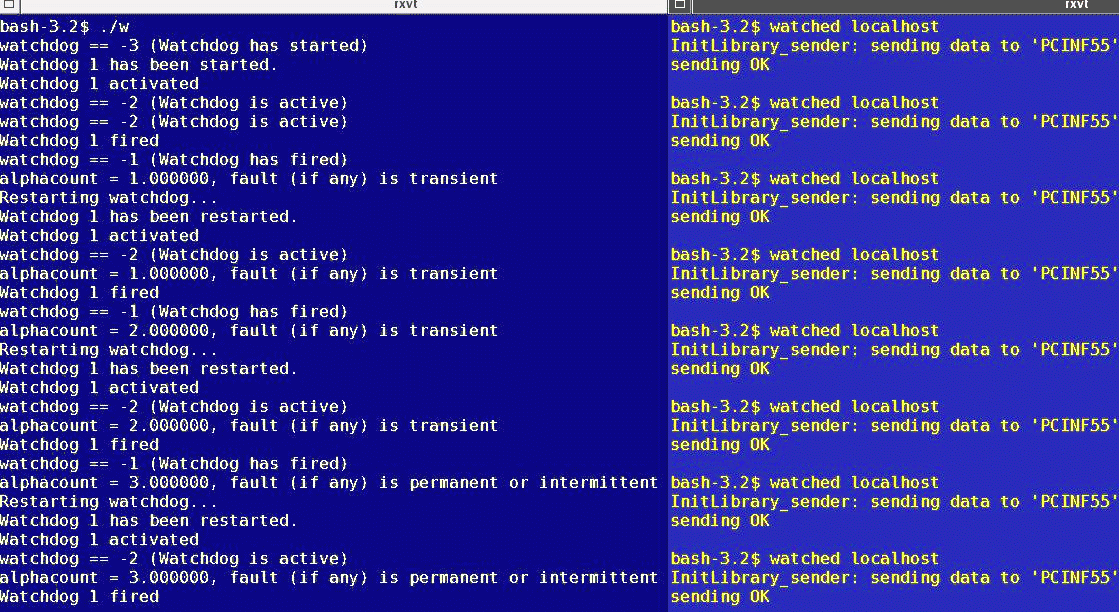}
\caption{A scenario involving a watchdog (left-hand window) 
and a watched task (right-hand). A permanent design fault
is repeatedly injected in the watched task.
As a consequence, the watchdog ``fires'' and an 
alpha-count variable is updated. The value of
that variable increases until it overcomes a threshold (3.0) and correspondingly the
fault is labeled as ``permanent or intermittent.''}
\label{f:alpha1}
\end{figure}

We have developed a prototypical version of this strategy (see Fig.~\ref{f:alpha1})
and we are now designing a full fledged version based on the cited ACCADA
and on an Alpha-count framework built with Apache Axis2~\cite{Axis2}
and MUSE~\cite{Muse}.

\subsubsection{Comparison with existing strategies.} 
Also in this case there exist strategies that postpone the
choice of the design pattern to execution time, though to the best of our
knowledge this has been done only with the design goal of achieving
performance improvements. A noteworthy example is FFTW,
a code generator for Fast Fourier Transforms that defines and assembles
(before compile time) blocks of C code that optimally solve FFT sub-problems
on a given machine~\cite{Fri04}.
Our strategy is clearly different in that
it focuses on dependability and makes use of a well-known count-and-threshold
mechanism.

%%%%%%%%%%%%%%%%%%%%%%%%%%%%%%%%%%%%%%%%%%%%%%%%%%%%%%%%%%%%%%%%%%%%%
\subsection{Assumptions Related to Dimensioning Replicated Resources}
%%%%%%%%%%%%%%%%%%%%%%%%%%%%%%%%%%%%%%%%%%%%%%%%%%%%%%%%%%%%%%%%%%%%%
As well known, a precise characterization of the amount of resources
necessary to deal with a certain situation is not always easy or even
possible to find out. In some cases, such amount is not to be considered
as a static value, fixed once and for all at design-time. Rather, it
should be modeled as a dynamic system, i.e. a variable changing over
time. When the situation to deal with is a threat to the quality
of a software service, then the common approach is to foresee
a certain amount of redundancy (time-, physical-, information-,
or design-redundancy). For instance, replication and voting can
be used to tolerate physical faults\footnote{Obviously
simple replication would not suffice to tolerate design faults,
in which case a design diversity scheme such as $N$-Version Programming
would be required.}. An important design problem is redundancy dimensioning.
Over-dimensioning redundancy or under-dimensioning it 
would respectively lead to either a waste of resources or failures.
Ideally the replication and voting scheme should work with a
number of replicas that closely follows the evolution of
the disturbance. In other words, the system should be aware
of changes in certain physical variables or at least of the
effect they are producing to its internal state. Not doing so---that
is, choosing once and for all a certain degree of redundancy---means
forcing the designer to take one assumption regarding the
expected range of disturbances. It also means that the system
will have a predetermined, necessary ``motion'' that will not be
affected by changes, however drastic or sudden. In other words,
the system will be a Boulding's \emph{Thermostat}. In what follow we describe
a strategy that can be used to enhance the Boulding category of a 
replication and voting scheme, thus avoiding a case of the \BS.

\begin{figure}
\centerline{\includegraphics[width=0.8\textwidth]{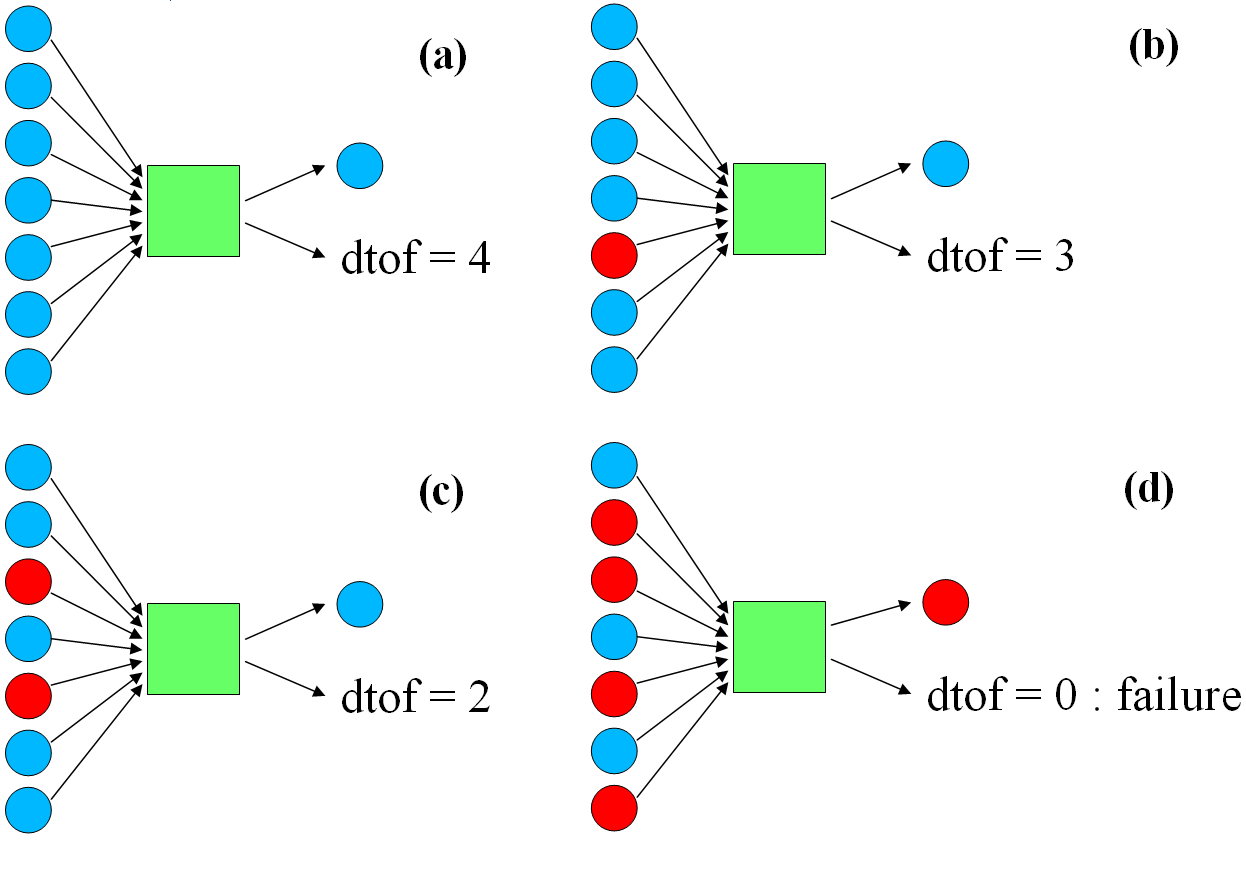}}
\caption{Distance-to-failure in a replication-and-voting scheme 
with 7 replicas. In (a), consensus is reached, which corresponds
to the farthest ``distance'' to failure. From (b) to (d), 
more and more votes dissent from the majority (red circles) and correspondingly
the distance shrinks. In (d), no majority can be found---thus,
failure is reached.}
\label{f:dtof}
\end{figure}

\begin{figure}
\centerline{\includegraphics[width=0.7\textwidth]{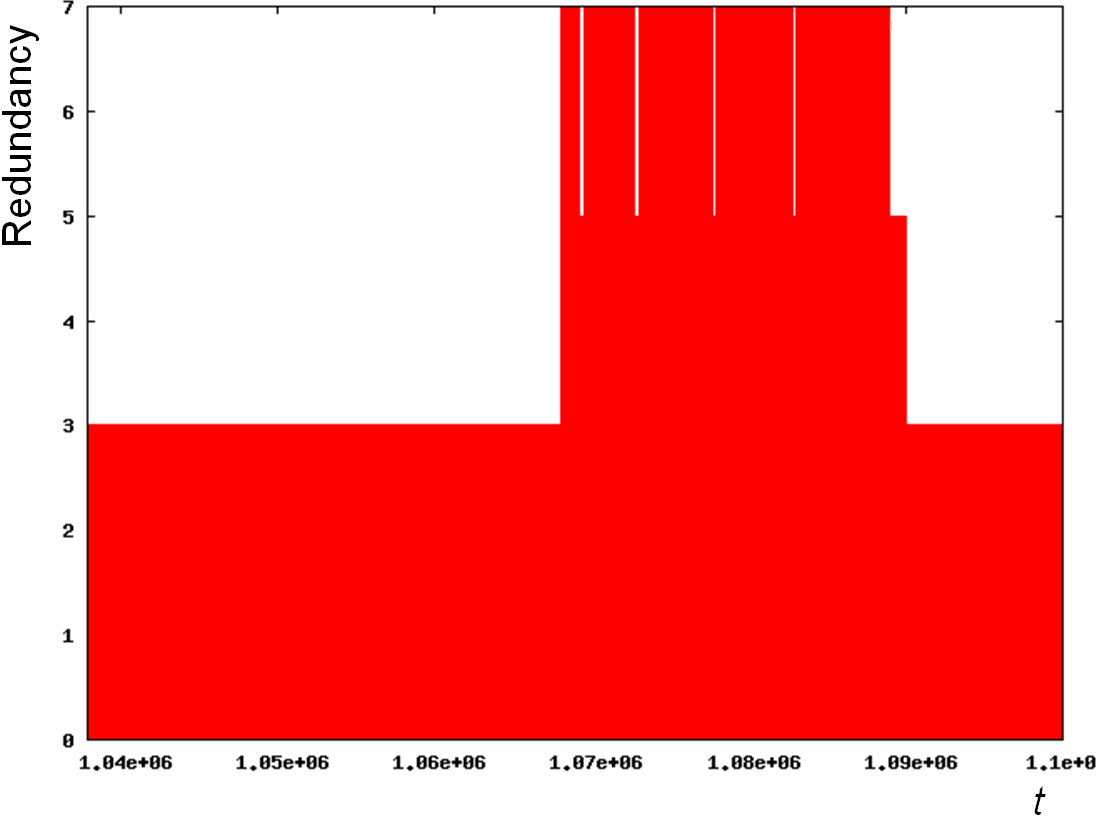}}
\caption{During a simulated experiment, faults are injected,
and consequently distance-to-failure decreases. This triggers an autonomic
adaptation of the degree of redundancy.}
\label{f:fromadidadr}
\end{figure}

The strategy we propose is to isolate redundancy management at
architectural level, and to use an autonomic computing scheme
to adjust it automatically.
In what follows we describe a possible implementation
for this run-time strategy.
\begin{itemize}
\item First, we assume that the replication-and-voting service
is available through an interface similar to the one of the
Voting Farm~\cite{DeDL98a}.
Such service sets up a so-called
``restoring organ''~\cite{John89a} after the user supplied the
number of replicas and
the method to replicate.
\item Secondly, we assume that the number
of replicas is not the result of
a fixed assumption but rather an initial value possibly subjected to revisions.
Revisions are triggered by secure messages that ask to raise or
lower the current number of replicas.
\item Third, we assume a middleware component such as our
Reflective Switchboards~\cite{DF10a} to be available. Such middleware
deducts and publishes a measure of the current environmental disturbances.
In our prototypical implementation, this is done by computing,
after each voting, the
``distance-to-failure'' (dtof), defined as 
\[
	\hbox{dtof}(n,m) = \lceil \frac n2 \rceil - m,
\]
where $n$ is the current number of replicas and $m$ is the amount
of votes
that differ from the majority, if any such majority exists.
If no majority can be found dtof returns 0.
%$\lceil \frac n2 \rceil$. As can be easily seen,
As can be easily seen,
dtof returns an integer in $[0, \lceil \frac n2 \rceil]$ that
represents how close we were to failure at the end of the last voting round.
The maximum distance is reached when there is full consensus among
the replicas. On the contrary the larger the dissent, the smaller
is the value returned by dtof, and the closer we are to
the failure of the voting scheme.
In other words, a large dissent (that is, small values
for dtof) is interpreted as 
a symptom that the current amount of redundancy employed is
not large enough.
Figure~\ref{f:dtof}
depicts some examples when the number of replicas is 7.
\item When dtof is critically low, the Reflective Switchboards
request the replication system to increase the number of redundant replicas.
\item When dtof is high for a certain amount of consecutive runs---1000 runs
in our experiments---a request to lower the number of replicas is issued.
Figure~\ref{f:fromadidadr} shows how redundancy varies in
correspondence of simulated environmental changes.
\end{itemize}

\begin{figure}[t]
%\centerline{\includegraphics[width=1.1\textwidth]{redundyn.png}}
\noindent% GNUPLOT: LaTeX picture
\setlength{\unitlength}{0.240900pt}
\ifx\plotpoint\undefined\newsavebox{\plotpoint}\fi
\sbox{\plotpoint}{\rule[-0.200pt]{0.400pt}{0.400pt}}%
\begin{picture}(1500,900)(0,0)
\sbox{\plotpoint}{\rule[-0.200pt]{0.400pt}{0.400pt}}%
\put(171.0,131.0){\rule[-0.200pt]{4.818pt}{0.400pt}}
\put(151,131){\makebox(0,0)[r]{ 3}}
\put(1430.0,131.0){\rule[-0.200pt]{4.818pt}{0.400pt}}
\put(171.0,255.0){\rule[-0.200pt]{4.818pt}{0.400pt}}
\put(151,255){\makebox(0,0)[r]{ 4}}
\put(1430.0,255.0){\rule[-0.200pt]{4.818pt}{0.400pt}}
\put(171.0,379.0){\rule[-0.200pt]{4.818pt}{0.400pt}}
\put(151,379){\makebox(0,0)[r]{ 5}}
\put(1430.0,379.0){\rule[-0.200pt]{4.818pt}{0.400pt}}
\put(171.0,504.0){\rule[-0.200pt]{4.818pt}{0.400pt}}
\put(151,504){\makebox(0,0)[r]{ 6}}
\put(1430.0,504.0){\rule[-0.200pt]{4.818pt}{0.400pt}}
\put(171.0,628.0){\rule[-0.200pt]{4.818pt}{0.400pt}}
\put(151,628){\makebox(0,0)[r]{ 7}}
\put(1430.0,628.0){\rule[-0.200pt]{4.818pt}{0.400pt}}
\put(171.0,752.0){\rule[-0.200pt]{4.818pt}{0.400pt}}
\put(151,752){\makebox(0,0)[r]{ 8}}
\put(1430.0,752.0){\rule[-0.200pt]{4.818pt}{0.400pt}}
\put(331.0,131.0){\rule[-0.200pt]{0.400pt}{4.818pt}}
\put(331,90){\makebox(0,0){3}}
\put(331.0,757.0){\rule[-0.200pt]{0.400pt}{4.818pt}}
\put(651.0,131.0){\rule[-0.200pt]{0.400pt}{4.818pt}}
\put(651,90){\makebox(0,0){5}}
\put(651.0,757.0){\rule[-0.200pt]{0.400pt}{4.818pt}}
\put(970.0,131.0){\rule[-0.200pt]{0.400pt}{4.818pt}}
\put(970,90){\makebox(0,0){7}}
\put(970.0,757.0){\rule[-0.200pt]{0.400pt}{4.818pt}}
\put(1290.0,131.0){\rule[-0.200pt]{0.400pt}{4.818pt}}
\put(1290,90){\makebox(0,0){9}}
\put(1290.0,757.0){\rule[-0.200pt]{0.400pt}{4.818pt}}
\put(171.0,131.0){\rule[-0.200pt]{0.400pt}{155.621pt}}
\put(171.0,131.0){\rule[-0.200pt]{308.111pt}{0.400pt}}
\put(1450.0,131.0){\rule[-0.200pt]{0.400pt}{155.621pt}}
\put(171.0,777.0){\rule[-0.200pt]{308.111pt}{0.400pt}}
\put(70,454){\makebox(0,0){$t$}}
\put(810,29){\makebox(0,0){Redundancy}}
\put(810,839){\makebox(0,0){Lifespan of assumption $a(r)$ : \{``Degree of employed redundancy is $r$''\}}}
\put(1290,249){\makebox(0,0){$log\,14648$}}
\put(651,194){\makebox(0,0){$log\,5631$}}
\put(970,279){\makebox(0,0){$log\,26534$}}
\put(331,697){\makebox(0,0){$log\,64953188$}}
\put(171.0,131.0){\rule[-0.200pt]{0.400pt}{144.058pt}}
\put(171.0,729.0){\rule[-0.200pt]{77.088pt}{0.400pt}}
\put(491.0,224.0){\rule[-0.200pt]{0.400pt}{121.654pt}}
\put(491.0,224.0){\rule[-0.200pt]{77.088pt}{0.400pt}}
\put(811.0,224.0){\rule[-0.200pt]{0.400pt}{20.236pt}}
\put(811.0,308.0){\rule[-0.200pt]{76.847pt}{0.400pt}}
\put(1130.0,276.0){\rule[-0.200pt]{0.400pt}{7.709pt}}
\put(1130.0,276.0){\rule[-0.200pt]{77.088pt}{0.400pt}}
\put(1450.0,131.0){\rule[-0.200pt]{0.400pt}{34.930pt}}
\put(171.0,131.0){\rule[-0.200pt]{0.400pt}{155.621pt}}
\put(171.0,131.0){\rule[-0.200pt]{308.111pt}{0.400pt}}
\put(1450.0,131.0){\rule[-0.200pt]{0.400pt}{155.621pt}}
\put(171.0,777.0){\rule[-0.200pt]{308.111pt}{0.400pt}}
\end{picture}
\caption{Histogram of the employed redundancy during an experiment
that lasted 65 million simulated 
time steps. For each degree of
redundancy $r$ (in this case $r\in \{3,5,7,9\}$) the graph displays the 
total amount of time steps the system adopted assumption
$a(r)$: \{``Degree of employed redundancy is $r$''\}.
A logarithmic scale is used
for time steps. Despite
fault injection, in the reported experiment the system spends
99.92798\% of its execution time
making use of the minimal degree of redundancy,
namely 3, without incurring in failures.}
\label{f:redundyn}
\end{figure}
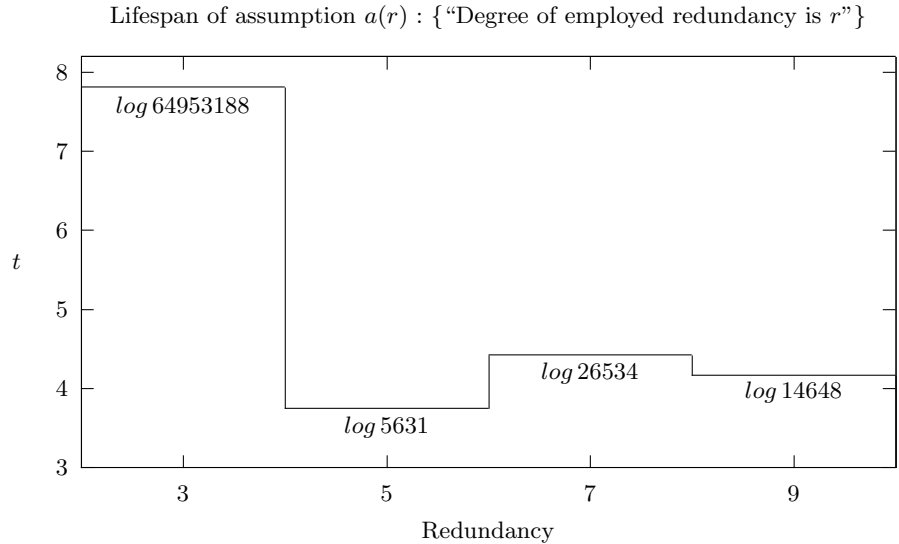

Function dtof is just one possible example of how to estimate the
chance of an impending assumption failure when dimensioning
redundant resources. 
Our experiments~\cite{DF10a} show that
even such a simplistic scheme allows most if not all dimensioning
assumption failures to be avoided. Despite heavy and diversified
fault injection, no clashes were observed during our experiments.
At the same time, as a side effect of assumption failure avoidance,
our autonomic scheme reduces the amount of redundant resources to be allocated and
managed. This can be seen for instance in
Fig.~\ref{f:redundyn} which plots in logarithmic scale the distribution of the
amount of redundancy employed by our scheme during one of our experiments.

The above strategy shows how \HS{} and \BS{} may be 
avoided---in a special case---by creating context-aware,
autonomically changing Horning Assumptions. In other
words, rather than postponing the decision of the value
to bind our assumption to, here we embedded our software
system in a very simple autonomic architecture that dynamically revises
dimensioning assumptions. The resulting system complies to
Boulding's categories of ``Cells'' and ``Plants'', i.e.
open software systems with a self-maintaining structure~\cite{Bou56}.

%%%%%%%%%%%%%%%%%%%%%%%%%%%%%%%%%%%%%%%%%%%%%%%
\section{Related Technologies}\label{s:reltech}
%%%%%%%%%%%%%%%%%%%%%%%%%%%%%%%%%%%%%%%%%%%%%%%
As mentioned in the introduction, several
conceptual and
practical tools are available to deal to some extent
with problems related to assumption failures. Such tools may be
situated in one or more of the following ``time stages'': design-time,
verification-time, compile-time, deployment-time, and run-time.
In what follows we briefly discuss some families of those
tools pointing out their relations with the subjects
treated in this paper.

\begin{description}
\item[Verification and validation activities\mbox{,}] i.e.,
checking and certification of compliance to specifications,
are a fundamental tool to verify and prove the absence of
some context clashes.
In particular re-qualification is an activity prescribed
each time a system (not necessarily a software system)
is ``relocated'' (e.g. reused, or ported);
or in case of replacement of some of its parts; or when
a system is connected to another system. 
We observe how, verification and validation being
in general off-line activities, assumptions
are matched against a reference context information
(the hypothized truth)
that might differ from the actual context---from ``real life'',
as it were.

Particularly interesting is the family of techniques known as
formal verification, which make use of formal (mathematical)
methods to assess a system's properties. Properties
are described through a formal specification.
Formal specification languages, such as the 
Z notation~\cite{z:baum95,z:bish90}, can be used for 
the non-ambiguous expression of software properties.
Compliant tools can then verify the validity of those
properties and detect cases of assumption failures.
Semantics~\cite{DBLP:journals/cacm/ShethVG06}
is another family of techniques that
aim at expressing and machine-verifying the meanings of
computing systems, their processing and environments.
\item[Unified Modeling Language] (UML) is the de-facto modeling
language for object-oriented software engineering.
A discussion of UML would lead
us astray, thus we shall just remark here how
UML provides means to describe
\begin{itemize}
\item the dependencies among the modeled software parts via component diagrams;
\item the mapping of software parts onto the target hardware and execution environment
      via deployment diagrams;
\item assorted knowledge, in the form of annotations;
\item rules and constraints, as for instance in the contexts and properties of
      the Object Constraint Language~\cite{OCL}.
\end{itemize}

UML and related tools situate themselves at design level though
can be used to generate implementation artifacts directly from
the models. By creating a stronger link between design-time
and other ``time stages'' such tools---when used correctly---make
it more difficult to incur in cases of the \HIS{} that are
due to model-vs.-code inconsistencies. We observe how the
produced artifacts are static entities that strictly follow
design-time rules; as such they are not able to self-adapt
so as to withstand faults or re-optimize in view of changed
conditions. In other words, those implementation artifacts
may suffer from the \BS.
\item[Design by Contract]~\cite{Mey92} is a design approach that systematically
deals with the mutual dependences of cooperating software components.
Depending on the context, any two software components may find
themselves in the role of a client and of a supplier of some service.
A well-defined ``contract'' formally specifies what are the obligations
and benefits of the two parties. This is expressed in terms of
pre-conditions, post-conditions, and invariants. Design by Contract
forces the designer to consider explicitly the mutual dependencies
and assumptions among correlated software components.
This facilitates assumption failures detection and---to some extent---treatment.
The concept of contracts has been recently successfully applied to security
of mobile applications~\cite{DMNS07,mw09.12}.
\item[Web Services standards] provide numerous examples
of specifications to expose, manage, and control
capabilities and features of web services architectures.
It is worth highlighting here a few of these standards:
\begin{description}
\item[WSDL] (Web Services Description Language) is an XML language
that allows a client to query and invoke the services exported by
any third-party web service on the Internet. This high degree
of flexibility exacerbates the problem of depending on third
party software components, i.e., software of unknown characteristics
and quality~\cite{Gre97}. The need to discipline this
potential chaos brought to a number of other specifications, such as
WS-Policy.
\item[WS-Policy] implements a sort of XML-based run-time version
of Design by Contract: using WS-Policy web service suppliers can
advertise their pre-conditions (expected requirements, e.g. 
related to security), post-conditions (expected state evolutions),
and invariants (expected stable states).
\item[WSDM] (Web Services Distributed Management)
and its two complementary specifications MUWS 
(Management Using Web Services) and MOWS (Management Of Web Services),
which respectively 
expose manageability capabilities and define a monitoring
and control model for Web Services resources. This 
allows for instance quality-of-service monitorings,
enforcing a service level agreement, or controlling a task.
\end{description}
\item[XML-based deployment descriptors] typical
of service-oriented and comp\-onent-oriented
middleware platforms such as J2EE or CORBA are
meant to reduce the complexity of deployment
especially in large-scale distributed systems.
Their main focus is clearly deployment-time.
Despite their widely recognized values,
some authors observe that they exhibit a
``semantic gap''~\cite{Sch06} between the design intent
and their verbose and visually dense syntax, which in practice
risks to conceal the very knowledge they are intended
to expose. This is probably not so relevant as the
exposed knowledge is meant to be reasoned upon by
machines.
\item[Introspection.]
  The idea of introspection
  is to gain access into the hidden software complexity, to inspect the
  black-box structure of programs, and to interpret their
  meaning through semantic processing, the same way the
  Semantic Web promises to accomplish with the data scattered through
  the Internet. Quoting~\cite{Intros}, ``introspection is a means
  that, when applied correctly, can help crack the code of a software
  and intercept the hidden and encapsulated meaning of the
  internals of a program''. Introspection
  is achieved e.g. by instrumenting software with data collectors producing
  information available in a form allowing semantic processing,
  such as RDF\cite{RDF}. This idea is being used in the Introspector
  project, which aims at instrumenting the GNU programming
  tool-chain so as to create a sort of semantic web of all
  software derived from those tools. The ultimate goal is
  very ambitious: ``To create a super large and extremely dense web of
  information about the outside world
  extracted automatically from computer language
  programs''~\cite{Intros}. This would allow the design of software able
  to reason about the dependability characteristics of other
  software. Tools based on introspection include:
  \begin{description}
    \item[GASTA] (Gcc Abstract Syntax Tree Analysis)~\cite{gasta}, which uses introspection to
     automatically annotate C code to analyze the presence of null pointer design faults), 
    \item[GCC-XML]~\cite{gccxml}, quite similar to GASTA, and
    \item[XOGASTAN] (XML-Oriented Gcc Abstract Syntax Tree ANalyzer)~\cite{ADM04},
     which uses the abstract syntax tree produced by the GNU compiler while processing a C file
     and translates it into XML. Another of the XOGASTAN tools can then read the XML file and analyze it.
  \end{description}
  In its current form introspection is an off-line technique working at code level.
\item[Aspect orientation] logically 
distinguishes a conventional language to encode the functional logics;
an aspect language to define 
specific interconnections among a program's basic functional units; and
a so-called aspect weaver, that is a program that composes a
software system from both the functional and the aspect logics.
Multiple aspect logics can be defined to address different
systemic cross-cutting concerns, e.g. enhancing dependability, minimizing
energy expenditure, or increasing performance. This has two
consequences particularly important for our treatise: the most
obvious one is that aspect oriented languages realize pliable
software that can be more easily maintained and adapted. Secondly,
aspects encode knowledge that regard specific ``viewpoints'', and
encourage the designers doing so.
As such, aspect orientation offers a conceptual and practical
framework to deal with the three syndromes of software development.
\item[Model Driven Engineering] (MDE) is a relatively new paradigm that
combines a number of the above approaches into a set of
conceptual and practical tools that address several shortcomings
of traditional methods of software development. In particular,
MDE recognizes that ``models alone are insufficient to
develop complex systems''~\cite{Sch06}.
Contrarily to
other approaches, which develop general ``languages'' to express
software models in an abstract way, MDE employs so-called
domain-specific modeling languages, which make use of semantics to
precisely characterize the relationships between concepts and their
associated constraints. The ability to express
domain-specific constraints and to apply model checking
allows 
several cases of assumption failures to be detected
early in the software life cycle.
Furthermore, MDE features
transformation engines and generators that synthesize
from the models 
various types of artifacts, e.g. source code and
XML deployment descriptions.
MDE systematically combines several existing technologies
and promises to become soon one of the most important ``tools''
to tame the ever growing complexity of software.
As remarked by Schmidt~\cite{Sch06}, the elegance
and the potential power of MDE brought about many expectations;
this notwithstanding,
scientific studies about
the true potential of MDE are still missing~\cite{Bez03,Bez05,Sch06}.
\end{description}

%%%%%%%%%%%%%%%%%%%%%%%%%%%%%%%%%%%%%%%%%%%%%%%%%%%%
\section{Lessons Learned and Vision}\label{s:vision}
%%%%%%%%%%%%%%%%%%%%%%%%%%%%%%%%%%%%%%%%%%%%%%%%%%%%

In previous section we discussed very concisely a few families of
approaches that can be effectively used to solve
some of the problems we introduced in this paper. Here we first summarize lessons
learned while doing so, which then brings us to our vision on future
approaches and architectures to deal effectively with assumption failures.

First of all, we can summarize that a number of
powerful techniques and tools exist or are in the course
of being honed that can effectively help dealing
with assumption failures. What is also quite apparent is
that each of them only tackles specific aspects of the
problem and takes a privileged view to it.

Our position in this context is that
we are still lacking methodologies and architectures to tackle this 
problem in its complex entirety. Fragmented views to this
very complex and entangled web are inherently ineffective
or at best sub-optimal. Missing one aspect means leaving a
backdoor open to the manifestations of the three syndromes introduced
in Sect.~\ref{s:hazards}.
In other words, a holistic approach is required. Taming
the complexity of software systems so as to reach true resilience
in the face of assumption failures
requires a unitary view to the whole of the ``time stages'' of
software development---what the General Systems Theory calls
a \emph{gestalt}~\cite{Bou56}. We believe
one such gestalt for software systems to be the concept of
assumption failure. As Boulding writes in the cited paper,
gestalts are ``of great value in directing research towards
the gaps which they reveal''---in the case at hand,
the gaps of each fragmented view offered by the
approaches mentioned in Sect.~\ref{s:reltech} to the 
problems discussed in this paper\footnote{In
  the cited paper Boulding applies this concept to the general system of disciplines 
  and theories: ``Each discipline corresponds to a certain segment of the empirical 
  world, and each develops theories which have particular applicability to its own 
  empirical segment. Physics, chemistry, biology, psychology, sociology, economics 
  and so on all carve out for themselves certain elements of the experience of man 
  and develop theories and patterns of activity (research) which yield satisfaction 
  in understanding, and which are appropriate to their special segments.'' 
  Gestalts, that is meta-theories of systems, ``might be of value in directing the 
  attention of theorists toward gaps in theoretical models, and might even be of 
  value in pointing towards methods of filling them.''}.
In a sense, most if not all of those approaches may be regarded
as the result of an attempt to divide and
conquer the complexity of software development by abstracting and
specializing (that is, reducing the scope of) methods, tools,
and approaches. This specialization
ends up in the ultimate case of the
Hidden Intelligence syndrome. A better approach would probably
be considering the unity of the design intent and using
a holistic, ``cross layered'' approach to share 
sensible knowledge unraveled in one layer and feed it back 
into the others.
We envision a general systems theory of
software development in
which the model,
compile-, deployment-, and run-time layers
feed one another with deductions and control ``knobs''.
The strategies discussed in this paper could provide
the designer with useful tools to arrange such cross-layering
processes.
This would allow knowledge
slipping from one layer to be still caught in another,
and knowledge gathered
in one layer to be fed back into others. As an example,
the strategy discussed in Sect.~\ref{s:pattern}
could feed an MDE tool whose deductions could in turn be published
or reified into a context-aware middleware such as our Reflective
Switchboards~\cite{DF10a}.

One way to achieve this could be to arrange a web of
cooperating reactive agents serving different software design
concerns (e.g. model-specific, deployment-specific, verification-specific,
execution-specific)
responding to external stimuli and
autonomically adjusting their internal state. Thus a design assumption
failure caught by a run-time detector should trigger a request for
adaptation at model level, and vice-versa. We believe that
such a holistic approach would realize
a more complete, unitary vision of a system's behavior and properties
with respect to the sum of the detached and fragmented views
available so far.

%%%%%%%%%%%%%%%%%%%%%%%%%%%%%%%%%%
\section{Conclusions}\label{s:end}
%%%%%%%%%%%%%%%%%%%%%%%%%%%%%%%%%%

Software systems are characterized by predefined assumptions about 
their intended platform, their internal state, and the environments they are 
meant to be deployed in. They are often closed, ``blind'' systems built from 
synchronous assumptions and designed so as to be plugged in immutable 
hardware systems and environments whose changes, idiosyncrasies, or 
fluctuations most of them deliberately ignore. We believe that this 
approach to software development is not valid anymore. Software ought to 
be designed and executed taking into account the inevitable occurrence 
of potentially significant and sudden changes or failures in their 
infrastructure and surrounding environments. By analyzing well-known 
software failures we identified three main threats to effective 
dependable software engineering, which we called the Hidden Intelligence 
syndrome, the Horning syndrome, and the Boulding syndrome. In this paper 
we expressed our thesis that services explicitly addressing those 
threats and requirements are an important ingredient towards truly 
resilient software architectures. For each of the above mentioned 
syndromes we also provided exemplary treatment strategies, which form 
the core of our current work in the adaptive-and-dependable software 
systems task force of the PATS research group at the University of 
Antwerp. The key idea is to provide the designer with the ability to 
formulate dynamic assumptions (assumption variables) whose boundings get 
postponed at a later, more appropriate, time: at compile time, when we 
are dealing with hardware platform assumptions for a stationary code; at 
deployment time, when the application can be assembled on that stage; 
and at run-time, e.g. when a change in the physical environment calls 
for adaptation to new environmental conditions. We believe that
an effective way to do this is by means of a web of
cooperating autonomic ``agents''
deducting and sharing knowledge, e.g. the type of faults being
experienced or the current values for properties regarding 
the hardware platform and the execution environment. 
We developed a number of these agents, e.g. Reflective
Switchboards~\cite{DF10a}, ACCADA~\cite{GuD10+}, and an Apache Axis2/MUSE
web service framework.
Our future steps include the design of a software architecture
for assumptions failure treatment based on the close cooperation
of those and other building blocks.

We conclude by observing how our research actually ``stands on the 
shoulders of giants'', as its conclusions closely follow those in the 
now classic 1956 paper of Kenneth Boulding on General Systems 
Theory~\cite{Bou56}: indeed, current software engineering practices 
often still produce systems belonging to Boulding's categories of 
``Clockworks'' and ``Thermostats''. The root assumption of such systems 
is their being closed-world, context-agnostic systems characterized by 
predefined assumptions about their platform, their internal state, and 
the environment they are meant to be deployed in, which makes them
fragile to change.
On the contrary, the unitary approach we envision,
based on the proposed role of gestalt for assumption failures,
would make it
possible to design and maintain actual open software systems with a 
self-maintaining structure (known as ``Cells'' and ``Plants'' according 
to Boulding's terminology) and pave the way to the design of
fully autonomically resilient software systems (Boulding's ``Beings'').

\subsection*{Acknowledments}
My gratitude goes to Jonas Buys, Ning Gui, and Hong Sun for providing
me with their helpful comments, suggestions, and valuable assistance, and
to Marc and Mieke Leeman for kindly providing me with the output from the
execution of the \texttt{lshw} command on several computers.

\bibliographystyle{splncs}
\bibliography{thesis}

\end{document}